# OVERVIEW DESCRIPTION OF THE GAMBIAN GABECE EDUCATIONAL DATA AND ASSOCIATED ALGORITHMS AND UNSUPERVIZED LEARNING PROCESS

OUSMANE SAINE [†], SOUMAILA DEMBÉLÉ [††], GANE SAMB LO [†††], MOHAMED CHEIKH HAIDARA [††††]

**Abstract.** As the first paper of a series of exploratory analysis and statistical investigation works on the Gambian *GABECE* data based on a variety of statistical tools, we wish to begin with a thorough unsupervised learning process through descriptive and exploratory methods. This will lead to a variety of discoveries and hypotheses that will direct future research works related to this data.

Ousman Saine [†]
University of The Gambia, Banjul, The Gambia
& Policy Planning, Analyses, research and Budgeting Directorate,
Ministry of Basic and secondary Education, Willy Thorpe Palace,
Banjul, The Gambia
ousmansaine@edugambia.gm

Dr Soumaila DEMBELE [††]
Université des Sciences Sociale et de Gestion de Bamako ( USSGB)
Faculté des Sciences Économiques et de Gestion (FSEG)
Email: soumidemlpot@gmail.com

[†††] Gane Samb Lo.
LERSTAD, Gaston Berger University, Saint-Louis, Sénégal (main affiliation).
LSTA, Pierre and Marie Curie University, Paris VI, France.
AUST - African University of Sciences and Technology, Abuja, Nigeria
gane-samb.lo@edu.ugb.sn, gslo@aust.edu.ng, ganesamblo@ganesamblo.net
Permanent address : 1178 Evanston Dr NW T3P 0J9,Calgary, Alberta, Canada.

Dr Mohamed Cheikh HAIDARA [††††]
Work Affiliation : Département de Mathématique , Université Cheikh Anta Diop (UCAD), Dakar, Senegal
Ecole Supérieure Polytechnique, Université Cheikh Anta Diop (UCAD), Dakar, Senegal
Research affiliation : LERSTAD, Université Gaston Berger (UGB), Saint-Louis, Senegal
Emails : chheikhh@yahoo.fr

1. INTRODUCTION

In this paper, the *GABECE* data is first presented and next statistically studied for the first time. We proceed to an unsupervised learning process. We will provide the description of the variables and consider the geographical and temporal sub-data. For each sub-data, different techniques will be applied : cleaning, outliers detection, frequency and related tables, spatial (geographical) or gender patterns, proportions of grades and evolution in time, etc.

A series of statistical analyzes will be undergone on the data by ourselves or other researchers. To prepare that, we wish to devote a note on the general presentation of the data and an unsupervised learning process with respect to the gender, the area and the time. Also **R** codes (R Core Team (2020) and Crawley (2009)) aiming at reading the data and getting the variables presented in Section 2 will be provided in Appendix II in page 21.

The rest of the paper is organized as follows. We will complete this introduction by the main findings in this paper and related recommendations. The data is fully described in Section 2. In Section 3, the *GABECE* data is compared with national statistics concerning population sizes of regions and genders to see whether or not access to education is fair in the Gambia. In section 4, the completeness of the the data is studied and it it will be shown that the data present un non-negligible number of incomplete data everywhere. In Section 5, a cleaning study is undergone by the detection of the outliers and it will appear that eliminating outliers, as usually recommended, would lead to a serious error since excellent students would simply ignored in the study. Instead of rejecting, we will propose regrouping the grades. In Section 6, we undergo a general investigation of the performances. We close the paper by conclusions in Section 7. Appendices are put after the bibliography. .
The main conclusions are summarized below.

(C1) Neither the regions nor the two genders are well represented in the educational GABECE data. The western regions, in general, have more students at grade 13 when compared with these real weighs in the country.

Although being more numerous in the country, females are under represented in the data. Even worse, the number of females in the data behaves chaotically where the number of males steadily increases with the year.

(R1) As a consequence, we strongly recommend to use sampling methods with sample sizes proportional to the real weighs of sub-groups in the global population.

(C2) There is a big number of incomplete data in all years, in all regions, and in each gender. After dropping them, we still have enough data to proceed to a significant analysis data.

(R2) On this side, we recommend statistical studies based on missing data estimation techniques. There is a lot to do on this, using nearest neighborhood, regression, classification, etc.

(C3) The statistical studies are about the grades in the core disciplines: Mathematics, Sciences, English, and SES. The grades range from 1 (excellent) to 9 (fail). Unfortunately, the level of students is not very fairly good since the fail grade is dominant almost everywhere. As well, there is a very few number of excellent students.

(R3) The result of that situation is that excellent grades 1, 2 or 3 are detected as outliers. So, in the statistical studies, such grades should be excluded. But, fortunately, a regrouping of all grades into three levels seems to be the best way to overcome that problem. Those three levels are: credit (regrouping credit grades 1 to 6), pass (regrouping 8 and 8) and fail.

The paper is destined to be a basis for a series of statistical studies on GABECE data and beyond on similar data for countries of the West African Examinations Council, whose The Gambia is member of, and on educational data in general.

## 2. Overall presentation of the *GABECE* data

The West African Examinations Council was established in 1952. This Council has the responsibility of ascertaining examinations required in the public interest in the English-speaking West African countries, comparable to those of equivalent examining authorities internationally, and to develop syllabuses, arranges and administers examinations and awards certificates. It also conducts a number international examinations.

The data we are going to use in this document is from the above aforementioned body. The data covers the period from the 2012 academic year to 2017. The mode of the GABECE examination is that each student is required to choose 7 subjects at least and 9 subjects at most. Out of the subjects taken, four (4) are compulsory and are called core subjects, which are English (ENG), Mathematics (MATH), Sciences (SCIENCES) and Social and Environmental Studies (SES).

Individual performance grants the student admission to Senior Secondary Education. The grading system in this result is categorized as follows : Credit (grades from 1 to 6), Pass (7 to 8), Fail (9). The result of a student is determined by the grades of the four core subjects plus the two best grades from the student's choice. The best result a student could have is the aggregated grade from 6 (six), one in all the core subjects, plus one in the two choices of the students and the worst result is aggregated to 54 with failure in all subjects with grade 9.

In this study, we focus on the core grades (G1), (G2), (G3) and (G4) corresponding to English, Mathematics, Sciences and *SES* respectively. Here are the conventions we will be using to know whether we use a grade for the whole population or not, for all the years or not, for both genders or not, or whether we use a grade for one particular gender, or a year or an area.

The corresponding variables of the core grades are denoted as $G1$, $G2$, $G3$ and $G4$ before the cleaning of the data and cover all areas, all time and all gender.

We also have the variables $year$ (from 2012 to 2017), gender (coded 1 for males and 2 for females), and region (coded for 1 to six regions).

The missing data, which correspond to absences of students or submission of blank sheets by student, represent 4.54% of the data (5792 out of 121838). The new variables become :

G1A, G2A, G3A, G4A, yearA, genderA, regionA.

When we restrict our analysis on a specific year for all gender, the variables are:

G1AY, G2AY, G3AY, G4AY, yearAY, genderAY, regionAY.

When we restrict our analysis on a specific region for all genders and years, the variables are :

G1AR, G2AR, G3AR, G4AR, yearAR, genderAR, regionAR.

When we restrict our analysis on a specific region **and** for a specific year, for all genders, the variables are :

G1AYR, G2AYR, G3AYR, G4AYR, yearAYR, genderAYR, regionAYR.

In each of the previous cases, the specified year and region are given by the variables *yearOfStudy* and *regionOfStudy*.

If we want to specify and fix a gender, we use the filtering codes : Variable[gender=="1"] for males (and Variable[gender=="2"] for females). For example, the variables G1AR[genderAR=="1"], G2AYR[genderAYR=="2"], G3A[genderA=="1"], respectively contain:

a) the grades for males in English for the region determined by the value of *regionOfStudy*,

b) the grades in Mathematics for females for the year and the region determined by the values assigned to *yearOfStudy* and to *regionOfStudy*,

c) the grades for males in Sciences for all regions and all years.

These notation are fixed for once and will be used in similar papers. More notation related to transformed variables may come later.

We post a complete R code in Appendix II in page 21 for the reading the data and forming the variable described above.

In this first release of studies on *GABECE*, we are going to give the Gambian national context of the data and compare the representation of specific population subgroups in the data. Next we describe the main characteristics of the data under the following points:

(1) facts and trends on the population;
(2) incomplete data;
(3) outliers;
(4) general trends of the performances;

| CO | 2012 | 2013 | 2014 | 2015 | 2016 | 2017 | Overall |
|---|---|---|---|---|---|---|---|
| Size | 20850 | 20388 | 20244 | 21987 | 21943 | 22218 | 127,630 |
| Male | 10672 | 10378 | 10553 | 11254 | 11617 | 12032 | 66,506 |
| Male (%) | 51.18 | 50.90 | 52.13 | 51.18 | 52.94 | 54.15 | 52.11 |
| Female | 10178 | 10010 | 9691 | 10733 | 10326 | 10186 | 61,112 |
| Female (%) | 48.82 | 49.10 | 47.90 | 48.82 | 47.06 | 45.85 | 47.89 |
| Region 1 | 6815 | 6717 | 6629 | 6843 | 6879 | 6832 | 40,715 |
| (%) | 32.69 | 32.95 | 32.75 | 31.12 | 31.35 | 30.75 | 31.90 |
| Region 2 | 8182 | 8001 | 8263 | 9752 | 9519 | 9984 | 53,701 |
| (%) | 39.24 | 39.24 | 40.82 | 44.35 | 43.38 | 44.94 | 42.08 |
| Region 3 | 2260 | 2149 | 2191 | 2090 | 2184 | 2012 | 12,886 |
| (%) | 10.83 | 10.54 | 10.82 | 9.50 | 9.95 | 9.06 | 10.10 |
| Region 4 | 1042 | 1079 | 847 | 812 | 854 | 917 | 5,551 |
| (%) | 5.00 | 5.29 | 4.18 | 3.69 | 3.89 | 4.12 | 4.35 |
| Region 5 | 1790 | 1603 | 1404 | 1467 | 1499 | 1388 | 9,151 |
| (%) | 8.59 | 7.86 | 6.94 | 6.67 | 6.83 | 6.25 | 7.17 |
| Region 6 | 761 | 839 | 910 | 1023 | 1008 | 1085 | 5626 |
| (%) | 3.65 | 4.12 | 4.50 | 4.65 | 4.59 | 4.88 | 4.41 |

TABLE 1. Number of Students by year, region and gender in absolute values and in percentages. Abbreviation: *CO*= counted Objects.

3. IN WHAT CONTEXT THE *GABECE* DATA REPRESENTS THE WHOLE POPULATION

The data concerns young students at age around 13 passing the grade 9 exam over the period 2012-2017. The Gambia has six administrative regions which are : (1) Kanifing (including Banjul), (2) Brikama, (3) Kerewan, (4) Mansa Konko, (5) Janjanbureh-Kuntaur, (6) Basse. That order corresponds to covering the country from East to West. In particular Basse is the extrem eastern area in the west from the Capital Banjul in Kanifing.

In total, 127,630 students took part to the *GABECE* exam and 5,792 missed at least one discipline exam or handed a blank copy.

The number of declared students by year, region and gender are given in Table 1. The number of Students by year, region and gender in absolute values and in percentages are posted in table 2.

| CO | 2012 | 2013 | 2014 | 2015 | 2016 | 20171 | All time |
|---|---|---|---|---|---|---|---|
| Region 1: Kanifing ||||||||
| Male | 3651 | 3627 | 3617 | 3698 | 3722 | 3856 | 22,171 |
| Female | 3164 | 3090 | 3012 | 3145 | 3157 | 2976 | 18,544 |
| Male (%) | 53.57 | 54.00 | 54.56 | 54.04 | 54.11 | 56.44 | 54.45 |
| Female (%) | 46.43 | 46.00 | 45.44 | 45.96 | 45.89 | 43.56 | 45.55 |
| Region 2: Brikama ||||||||
| Male | 4002 | 3897 | 4139 | 4896 | 4968 | 5303 | 27,205 |
| Female | 4180 | 4104 | 4124 | 4856 | 4551 | 4681 | 26,469 |
| Male (%) | 48.91 | 48.71 | 50.09 | 50.21 | 52.19 | 53.11 | 50.66 |
| Female (%) | 51.09 | 51.29 | 49.91 | 49.79 | 47.81 | 46.89 | 49.34 |
| Region 3: Mansa Konko ||||||||
| Male | 1171 | 1065 | 1161 | 1019 | 1115 | 1052 | 6,583 |
| Female | 1089 | 1084 | 1030 | 1071 | 1069 | 960 | 6,303 |
| Male (%) | 51.81 | 49.56 | 52.99 | 48.76 | 51.05 | 52.29 | 51.09 |
| Female (%) | 48.19 | 50.44 | 47.01 | 51.24 | 48.95 | 47.71 | 48.91 |
| Region 4: Janjanbureh ||||||||
| Male | 526 | 516 | 420 | 384 | 449 | 483 | 2,778 |
| Female | 516 | 563 | 427 | 428 | 405 | 434 | 2,773 |
| Male (%) | 50.48 | 47.82 | 49.59 | 47.29 | 52.58 | 52.67 | 50.05 |
| Female (%) | 49.52 | 52.18 | 50.41 | 52.71 | 47.42 | 47.33 | 49.95 |
| Region 5: Kerewan ||||||||
| Male | 1048 | 920 | 804 | 799 | 872 | 832 | 5,273 |
| Female | 742 | 683 | 600 | 668 | 627 | 556 | 3,876 |
| Male (%) | 58.55 | 57.39 | 57.26 | 54.46 | 58.17 | 59.94 | 57.64 |
| Female (%) | 41.45 | 42.61 | 42.74 | 45.54 | 41.83 | 40.06 | 42.36 |
| Region 6: Basse ||||||||
| Male | 274 | 353 | 412 | 458 | 491 | 506 | 2.494 |
| Female | 487 | 486 | 498 | 565 | 517 | 579 | 3.132 |
| Male (%) | 36.01 | 42.07 | 45.27 | 44.77 | 48.71 | 46.64 | 44.33 |
| Female (%) | 63.99 | 57.93 | 54.73 | 55.23 | 51.29 | 53.36 | 55.67 |

TABLE 2. Number of Students by year, region and gender in absolute values and in percentages. *CO*: counted Objects.

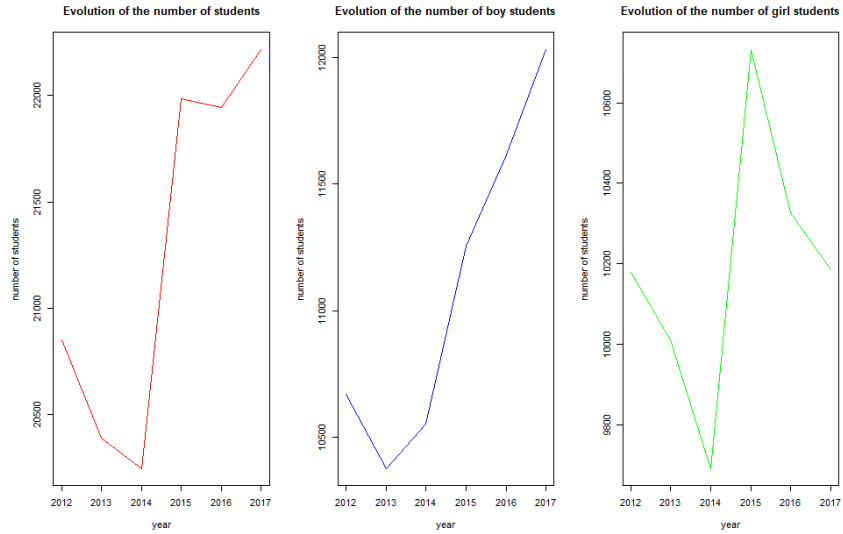

Figure 1. Evolution of number of students over the years (globally at left, for boys at the center, for girls at right)

We may have expected that the number of pupils is increasing from year 2012 to 2017. However, as shown in Fig. 1, we see drops in 2013 and 2014 for the total number of students before a tendency of increase which also is broken in 2016. For boys, the positive trend is remarkable after a drop in 2013. The evolution of the number of female is somehow chaotic, suggesting a lack of a successful gender policy to get girls into the educational system or to keep them in.

We will compare the distribution of the regions and the gender with the national statistics. In Table 3, we have the distribution of the population over eight (8) regions (Banjul, Kanifing, Brikama, Kerewan, Kuntaur, Janjanbureh, Basse). In Fig. 2, we have the map of the Gambia. The country, which is entirely surrounded by Sén'egal except at the Atlantic ocean coast, seems to be caught by a nearly rectangle horizontally ranging from East (Atlantic Ocean) to West in the interior of SENEGAL.

In the *GABECE* data, the regions of Banjul and Kanifing are regrouped in one region named after the later (Kanifing) and the regions of Janjanbureh and Kauntur are regrouped in one region named after the first. In Table 4, we finally have the distribution in percentages of the population over the six geographical areas used in the *GABECE* data. That table also displays

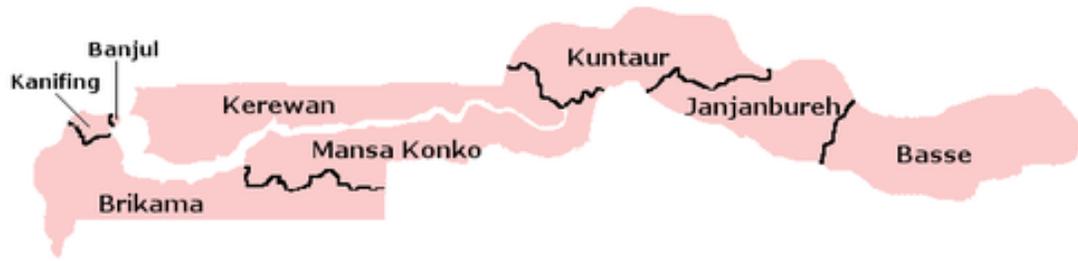

FIGURE 2. Map of the Gambia. The region of Janjanbureh and Kuntaur are regrouped and named after Kuntaur

| Regions | Area (km2) | Pop.C.03 | Pop.C.13 | Capital | N.D |
|---|---|---|---|---|---|
| Banjul (C) | 12.2 | 35,061 | 31,301 | Banjul | 3 |
| Kanifing | 75.6 | 322,735 | 382,096 | Kanifing | 1 |
| Brikama | 1,764.3 | 389,594 | 699,704 | Brikama | 9 |
| Mansa Konko | 1,628.0 | 72,167 | 82,381 | Mansakonko | 6 |
| Kerewan | 2,255.5 | 172,835 | 221,054 | Kerewan | 7 |
| Kuntaur | 1,466.5 | 78,491 | 99,108 | Kuntaur | 5 |
| Janjanbureh | 1,427.8 | 107,212 | 126,910 | Janjanbureh | 5 |
| Basse | 2,069.5 | 182,586 | 239,916 | Basse | 7 |
| Total Gambia | 10,689 | 1,360,681 | 1,882,450 | Banjul | 43 |

TABLE 3. Population by region in 2003 and 2013 (provisional). Abbreviations: pop.C.03 = population Census 2003; pop.C.13 = population Census 2013; (C) Capital of the country; N.D= number of districts. Source: Wikipedia (2020)

the areas in $km^2$ of each region.

Fig. 3 shows the weighs of the regions in the country. We can see that the Region 1 has the greatest population density followed by the three extreme western areas.

As to the gender repartition, females are slightly more dominant of males over the study period as shown in Fig. 4

| Num | Regions | Area (km2) | Pop.C.03 | % | Pop.C.13 | % |
|---|---|---|---|---|---|---|
| 1 | Kanifing | 87.4 | 357796 | 26.30% | 413397 | 21.96% |
| 2 | Brikama | 1764.3 | 389594 | 28.63% | 699704 | 37.17% |
| 3 | Mansa Konko | 1628 | 72167 | 5.30% | 82381 | 4.38% |
| 4 | Kerewan | 2255.5 | 172835 | 12.70% | 221054 | 11.74% |
| 5 | Janjanburey | 2894.3 | 185703 | 13.65% | 226018 | 12.01% |
| 6 | Basse | 2069.5 | 182586 | 13.42% | 239916 | 12.74% |
|  | Total Gambia | 10689 | 1360681 | 100.00% | 1882450 | 100.00% |

TABLE 4. Populations of regions in % in Census 2003 and census 2013

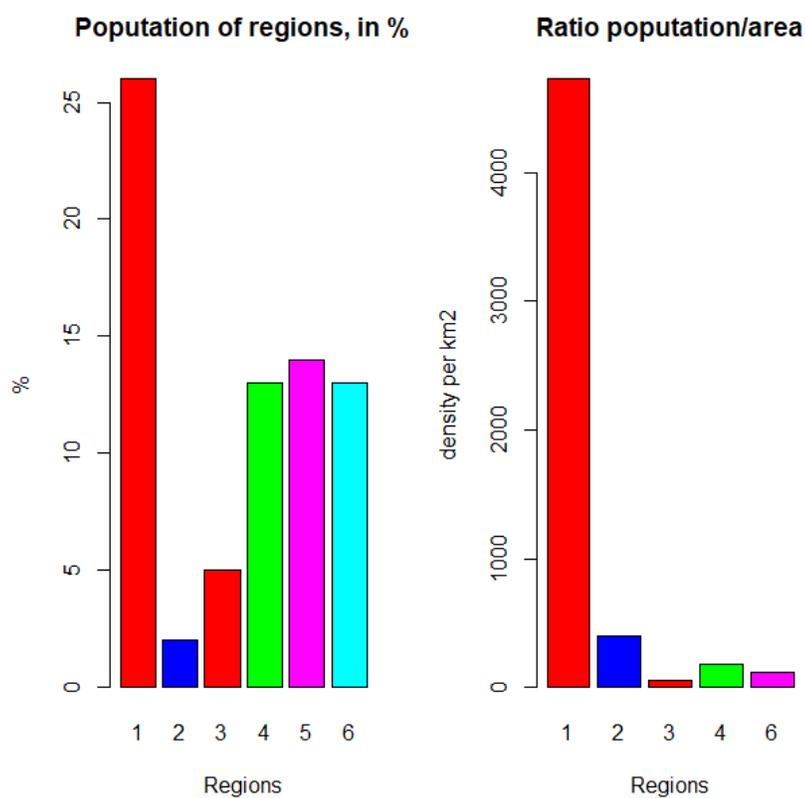

FIGURE 3. Populations in % of regions nationally (left) and in the data (right)

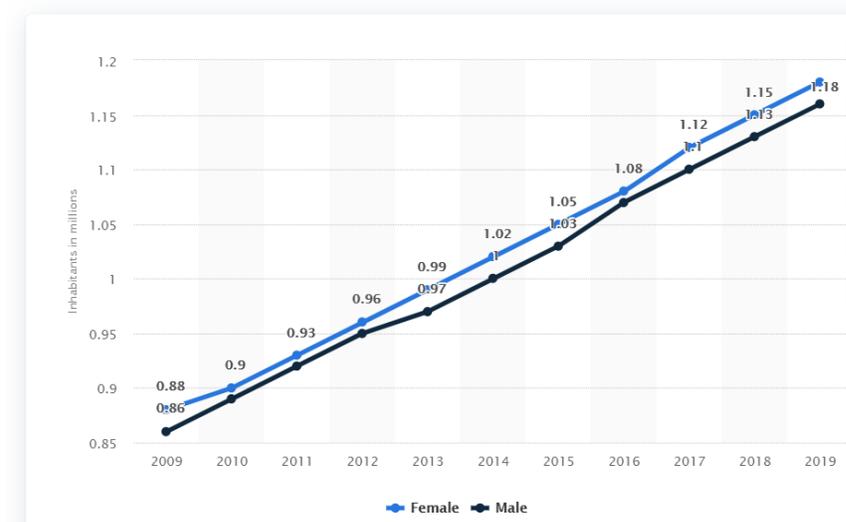

FIGURE 4. Domination of the female gender over of male gender over years. Source: Statistica.com (2020)

Now, let us see how the distribution of the *GABECE* compare with the national data with respect to the region and the gender.

**a. The regions are not equally represented**.

The comparison of the percentages in Table 3 with those in Table 5 are summarized in Table 5. The plot of the ratios of region percentage in the data in the region percentage in the GABECE data, in Fig. 5, shows that the three western region are over represented by far in the education data, the extreme western is not almost equally represented and the three middle regions are seriously under represented. Indeed an almost equal presence of the regions in the data should have led to points very near the bisector line. The more a point is above the bisector line, the more the corresponding region is over represented and the more a point is below the bisector line, the more the corresponding region is under represented.

**b. The female gender is poorly represented in the education data**.

From Tables 1 and 2 and Fig. 5, it very clear that female Gambian are badly represented in the education system although they constitute the majority in the period of the study. At the exception of the Basse area, more boys arrive at Level 13 that girls from 2014 to 2018. The biggest

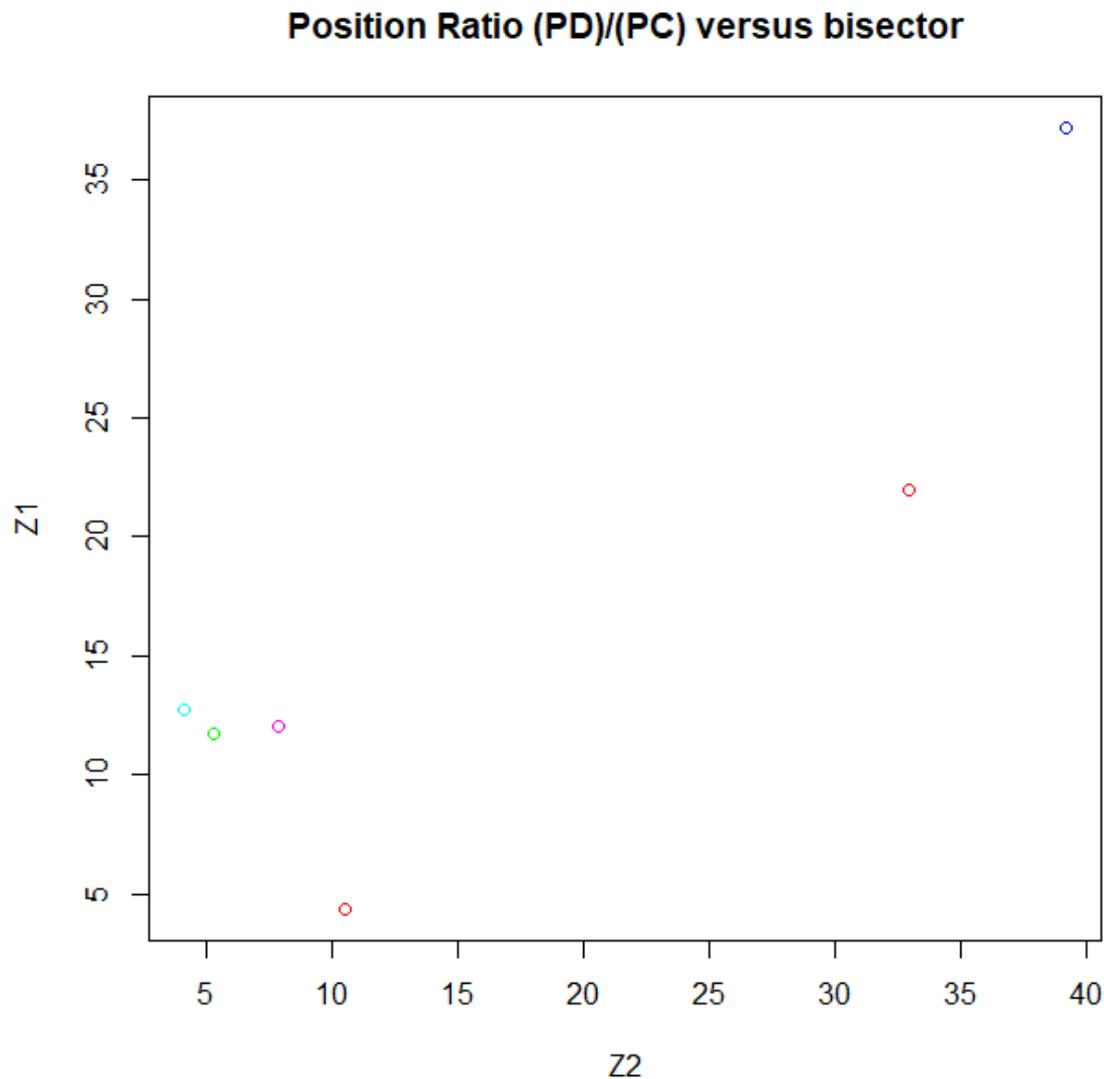

FIGURE 5. Scatterplot population of regions nationally and population of regions in the data

gaps happen in Kerewan.

How to explain why the contrary happens in the rural zone of Basse in opposite to Kerewan area which is adjacent to it? Local studies should be undergone to study such microscopic facts.

| Num | Regions | Presence in the Gambia | Presence in the data |
|---|---|---|---|
| 1 | Kanifing | 21.96% | 32.95% |
| 2 | Brikama | 37.17% | 39.17% |
| 3 | Mansa Konko | 4.38% | 10.54% |
| 4 | Kerewan | 11.74% | 5.29% |
| 5 | Janjanburey-kuntaur | 12.01% | 7.83% |
| 6 | Basse | 12.74% | 4.14% |

TABLE 5. Population of the regions in % nationally and in the dat

In Brikama, for 2012 and 2013, girls were well enrolled in the system according to their majority but the tendency came short to be persistent.

**Conclusion 1**. The *GABECE* data does not really represent the national statistics. Some regions are over represented (as the first region from East). The others are under represented with Basse being not very under present in the education data.

As well, the majority of the female population is not materialized in the eduction data. This, surely, is a sign of a poor gender equity policy to be solved.

**Recommendation 1**. In the statistical study, samplings should be used to extract from the data samples corresponding to national ratios for each region and gender to expect to have a fair comparison between of regions and of genders.

## 4. INCOMPLETE DATA

The data presents a high number of incomplete data. The missing data concerns the grades G1, G2, G3, G4. For a grade, a missing data (-1 or -2) is assigned if the student did not attend the exam of that grade or he/she handed a blank copy. We only give here the percentages of missing data by region and year. We see that 2.5% to roughly 8% of the data are missing data.

| **Regions** | 1 | 2 | 3 | 4 | 5 | 6 |
|---|---|---|---|---|---|---|
| **Data** | 1550 | 2684 | 519 | 246 | 378 | 415 |
| **Data %** | 3.8069 | 4.9980 | 4.0276 | 4.4316 | 4.1306 | 7.3764 |

TABLE 6. incomplete data by region

| **Years** | 2012 | 2013 | 2014 | 2015 | 2016 | 2017 |
|---|---|---|---|---|---|---|
| **Data** | 652 | 589 | 519 | 1138 | 1709 | 1185 |
| **Data %** | 3.1270 | 2.8889 | 2.5637 | 5.1757 | 7.7883 | 5.3335 |

TABLE 7. incomplete data by year

## 5. OUTLIERS

As explained in the introduction, the grades range from very excellent (1) to fail (9), with credits between 1 to 6. Grades (7) and (8) are for passing students with very average levels.

We will see later that the good students with grades (1), (2), (3) are very rare. They are so few that they seem to be outliers, that is data that can be taken as exceptional or accidental.

At the same times a bulk of the grades range over $\{7, 8, 9\}$.

The consequence is that outlier tests do detect students with credits as exceptional data, that should be studied apart. Here we use the simple Wilks tests for outliers which works as follows:

- Compute the first quartile $Q_1$, the second first quartile $Q_2 = Me$ which is the median and the third quartile $Q_3$
- compute the interquartile distance $D = Q_3 - Q_2$
- Compute the bounds $d_1 = Me - 1.5D$ and $d_2 = Me + 1.5D$.
- Declare outlier at left (OL) all elements below $d_1$ and outliers at right (OR) all observations beyond $d_2$

We applied that simple rule to each grade G1, G2, G3 and G4. The percentages of outliers at left (OL) and outliers at right (OR) are given by grade in Tables 8 - 11. In each table, The percentages of outliers at left (OL) and outliers at right (OR) are posted for every region. When we consider data by regions, the outliers are denoted by (OLr) [at left] and (ORr) [at left]. When data are restricted to regions and to a gender, outliers are denoted by (OLrf) [left] and (ORrf) [right] for females, (OLrm) [left] and (ORrm) [right]

| Regions | 1       | 2       | 3       | 4       | 5       | 6       |
|---------|---------|---------|---------|---------|---------|---------|
| **OLr**  | 12.0388 | 14.2089 | 24.4279 | 23.9585 | 18.1123 | 20.6486 |
| **ORr**  | 0       | 0       | 0       | 0       | 0       | 0       |
| **OLrf** | 10.6113 | 10.0593 | 11.3523 | 11.8554 | 9.1617  | 24.2392 |
| **ORrf** | 0       | 0       | 0       | 0       | 0       | 0       |
| **OLrm** | 13.2076 | 13.3001 | 20.5589 | 16.9184 | 20.2336 | 22.8978 |
| **ORrm** | 0       | 0       | 0       | 0       | 0       | 0       |

TABLE 8. Outliers in English grade by region and year

| Regions | 1       | 2       | 3       | 4       | 5       | 6       |
|---------|---------|---------|---------|---------|---------|---------|
| **OLr**  | 23.4750 | 18.4683 | 16.9887 | 21.2629 | 14.9777 | 23.1817 |
| **ORr**  | 0       | 0       | 0       | 0       | 0       | 0       |
| **OLrf** | 24.5746 | 20.3071 | 18.5780 | 9.7478  | 18.0010 | 13.9349 |
| **ORrf** | 0       | 0       | 0       | 0       | 0       | 0       |
| **OLrm** | 22.5746 | 16.7094 | 15.4745 | 17.1072 | 24.5693 | 19.3186 |
| **ORrm** | 0       | 0       | 0       | 0       | 0       | 0       |

TABLE 9. Outliers in Mathematics grades by region and year

for males.

The percentage of outliers range roughly between 3% and 25%. The biggest values are related to male students.

The results confirm what has been said before.

**Conclusion 2**. Excellent grades are so rare that they have been declared outliers at left. Weak grades so very common that none of them is declared outlier at right.

**Recommendation 2**. There is no statistical interest to study the excellent grades. So it is highly recommended to regroup the grades into meaningful classes. For example, we may regroup them into two class: Pass (including all grades not equal to 9) and Fail for grade 9; or into three class: credit (from 1 to 6), Pass (7 and 8) and Fail (9).

| Regions | 1 | 2 | 3 | 4 | 5 | 6 |
|---|---|---|---|---|---|---|
| **OLr** | 14.3138 | 10.7826 | 23.8861 | 10.2356 | 18.4429 | 18.6336 |
| **ORr** | 0 | 0 | 0 | 0 | 0 | 0 |
| **OLrf** | 15.0862 | 12.6854 | 10.8883 | 14.1512 | 9.0053 | 22.3374 |
| **ORrf** | 0 | 0 | 0 | 0 | 0 | 0 |
| **OLrm** | 13.6813 | 23.3288 | 19.7852 | 18.2779 | 23.2627 | 21.5179 |
| **ORrm** | 0 | 0 | 0 | 0 | 0 | 0 |

TABLE 10. Outliers in Sciences grades by region and year

| Regions | 1 | 2 | 3 | 4 | 5 | 6 |
|---|---|---|---|---|---|---|
| **OLr** | 4.6572 | 14.1776 | 12.0805 | 2.7144 | 12.1395 | 12.9533 |
| **ORr** | 0 | 0 | 0 | 0 | 0 | 0 |
| **OLrf** | 0 | 3.4279 | 2.9830 | 0 | 2.5523 | 13.0013 |
| **ORrf** | 0 | 0 | 0 | 0 | 0 | 0 |
| **OLrm** | 4.4489 | 14.9146 | 11.6374 | 11.8957 | 23.0647 | 24.5364 |
| **ORrm** | 0 | 0 | 0 | 0 | 0 | 0 |

TABLE 11. Outliers in SES grades by region and year

## 6. General analysis of performances

We are giving some trends of the performances. More insight studies will be made in up coming papers.

Table 12 gives the percentages of grades (1 to 9) by discipline and year. We can see the following conclusions:

(C1) The fail grade is overwhelmingly dominant in Mathematics, Sciences and English disciplines. Even for SES which can be thought less tough that the three others, fail rates are significant and turn around 50%.

(C2) The excellency credit grade (1) is extremely rare and its rate does not exceed 2% at the exception of the disciplines SES for which it is still low and less than 5%.

(C3) The other credit grades are still low enough to be studied apart.

| disp/year | cr 1 | cr 2 | cr 3 | ce 4 | cr 5 | cr 6 | pass 1 | pass 2 | fail |
|---|---|---|---|---|---|---|---|---|---|
| G1A 2012 | 1.75 | 2.13 | 6.62 | 4.44 | 3.73 | 8.76 | 6.94 | 9.66 | 55.97 |
| G2A 2012 | 0.54 | 0.72 | 2.20 | 1.59 | 1.82 | 5.69 | 8.12 | 12.99 | 66.33 |
| G3A 2012 | 0.43 | 0.60 | 2.91 | 2.19 | 3.00 | 7.85 | 4.94 | 7.88 | 70.19 |
| G4A 2012 | 3.10 | 2.89 | 9.52 | 4.41 | 5.15 | 12.60 | 5.50 | 10.57 | 46.27 |
| G1A 2013 | 0.56 | 0.81 | 3.23 | 2.66 | 2.45 | 6.72 | 7.61 | 10.60 | 65.35 |
| G2A 2013 | 0.67 | 1.03 | 3.73 | 2.34 | 2.76 | 7.17 | 6.84 | 10.14 | 65.31 |
| G3A 2013 | 0.31 | 0.53 | 2.46 | 1.75 | 1.74 | 8.33 | 6.65 | 11.24 | 66.98 |
| G4A 2013 | 2.10 | 2.48 | 8.16 | 3.89 | 4.65 | 11.30 | 7.09 | 10.20 | 50.12 |
| G1A 2014 | 0.19 | 0.60 | 2.93 | 2.48 | 2.34 | 8.97 | 9.67 | 12.48 | 60.35 |
| G2A 2014 | 0.60 | 0.61 | 2.15 | 0.76 | 1.53 | 5.28 | 6.02 | 10.33 | 72.71 |
| G3A 2014 | 1.80 | 1.12 | 3.90 | 2.62 | 2.40 | 8.15 | 8.61 | 14.68 | 56.74 |
| G4A 2014 | 2.50 | 2.54 | 9.22 | 4.55 | 5.11 | 12.29 | 5.42 | 10.40 | 47.96 |
| G1A 2015 | 1.27 | 1.53 | 6.27 | 4.06 | 3.93 | 11.14 | 9.69 | 11.37 | 50.74 |
| G2A 2015 | 0.74 | 0.69 | 2.59 | 1.39 | 1.82 | 5.21 | 2.49 | 15.83 | 69.25 |
| G3A 2015 | 0.29 | 0.58 | 2.83 | 1.32 | 2.36 | 8.07 | 9.59 | 14.60 | 60.38 |
| G4A 2015 | 2.53 | 2.46 | 7.18 | 4.93 | 3.38 | 11.46 | 7.46 | 10.87 | 49.71 |
| G1A 2016 | 0.32 | 1.00 | 5.90 | 3.18 | 5.33 | 10.07 | 7.94 | 11.08 | 55.17 |
| G1A 2016 | 0.85 | 0.86 | 2.98 | 1.50 | 1.48 | 5.77 | 7.15 | 13.70 | 65.69 |
| G2A 2016 | 1.45 | 1.36 | 4.19 | 2.90 | 2.18 | 8.50 | 8.01 | 10.90 | 60.51 |
| G3A 2016 | 3.18 | 3.56 | 9.39 | 5.36 | 5.27 | 13.48 | 6.06 | 8.68 | 45.03 |
| G4A 2017 | 0.45 | 1.11 | 6.13 | 4.21 | 10.34 | 7.59 | 8.00 | 9.14 | 53.03 |
| G2A 2017 | 0.69 | 0.82 | 2.68 | 1.14 | 4.04 | 3.54 | 6.71 | 12.74 | 67.64 |
| G3A 2017 | 1.82 | 1.56 | 5.21 | 2.90 | 7.69 | 5.92 | 7.54 | 8.24 | 59.12 |
| G4A 2017 | 4.88 | 4.03 | 10.91 | 4.93 | 11.40 | 7.16 | 7.07 | 7.74 | 41.88 |

TABLE 12. Percentages of grades (1 to 9) by disciplines and by year

# 7. CONCLUSIONS

This paper is destined to be the basis of a series of more insightful papers using a variety of more or less deep statistical techniques. At least, the data is entirely described. The notations are posed. A general algorithm allowing to get all relevant variables and their local restrictions (in time (year), in area (region), in gender (female, male)).

First statistical trends are given, concerning incomplete data, outliers, fairness of representation of sub-groups in the educational data, some general evolutions of grades in the time. These statistical facts already direct to interesting leads and suggest hypotheses to be validated or not.

| garde/year | cr1-4 | cr1-5 | cr1-6 | pass | fail |
|---|---|---|---|---|---|
| G1A 2012 | 14.94 | 18.67 | 27.43 | 44.03 | 55.97 |
| G2A 2012 | 5.05 | 6.87 | 12.56 | 33.67 | 66.33 |
| G3A 2012 | 6.13 | 9.13 | 16.98 | 29.8 | 70.19 |
| G4A 2012 | 19.92 | 25.07 | 37.67 | 53.74 | 46.27 |
| G1A 2013 | 7.26 | 9.71 | 16.43 | 34.64 | 65.35 |
| G2A 2013 | 7.77 | 10.53 | 17.7 | 34.68 | 65.31 |
| G3A 2013 | 5.05 | 6.79 | 15.12 | 33.01 | 66.98 |
| G4A 2013 | 16.63 | 21.28 | 32.58 | 49.87 | 50.12 |
| G1A 2014 | 6.2 | 8.54 | 17.51 | 39.66 | 60.35 |
| G2A 2014 | 4.12 | 5.65 | 10.93 | 27.28 | 72.71 |
| G3A 2014 | 9.44 | 11.84 | 19.99 | 43.28 | 56.74 |
| G4A 2014 | 18.81 | 23.92 | 36.21 | 52.03 | 47.96 |
| G1A 2015 | 13.13 | 17.06 | 28.2 | 49.26 | 50.74 |
| G2A 2015 | 5.41 | 7.23 | 12.44 | 30.76 | 69.25 |
| G3A 2015 | 5.02 | 7.38 | 15.45 | 39.64 | 60.38 |
| G4A 2015 | 17.1 | 20.48 | 31.94 | 50.27 | 49.71 |
| G1A 2016 | 10.4 | 15.73 | 25.8 | 44.82 | 55.17 |
| G1A 2016 | 6.19 | 7.67 | 13.44 | 34.29 | 65.69 |
| G2A 2016 | 9.9 | 12.08 | 20.58 | 39.49 | 60.51 |
| G3A 2016 | 21.49 | 26.76 | 40.24 | 54.98 | 45.03 |
| G4A 2017 | 11.9 | 22.24 | 29.83 | 46.97 | 53.03 |
| G2A 2017 | 5.33 | 9.37 | 12.91 | 32.36 | 67.64 |
| G3A 2017 | 11.49 | 19.18 | 25.1 | 40.88 | 59.12 |
| G4A 2017 | 24.75 | 36.15 | 43.31 | 58.12 | 41.88 |

TABLE 13. Percentages of regrouped grades (credit, pass, fail) by discipline and by year

**Acknowledgement** The authors wish to thank the Ministry .... that allowed to use the data.

# Appendix : R codes

See Crawley (2009) for an overall review on R software.

```
#The data are in the directory gabece, which is
#of a subdirectory where is this script.

#The directory should contain the following files relation to
#the whole population of students, having the same lenghth

#1 - mesAppurements.txt : contains the variable okData with two values
#"ok" for complete data "ko" for incomplete data.\\
#2 - the original grades G1, G2, G2, and G1.
#3 - the covariates year.txt, gender.txt and region.txt

appu <- read.table("gabece/mesAppurements.txt", h=TRUE)
attach(appu)
table(okData)

Gender <- read.table("gabece/gender.txt", h=TRUE)
attach(Gender)
gender
table(gender)

Year <- read.table("gabece/year.txt", h=TRUE)
attach(Year)
table(year)

Region <- read.table("gabece/region.txt", h=TRUE)
attach(Region)
table(region)

English <- read.table("gabece/G1.txt", h=TRUE)
attach(English)
table(G1)

Maths <- read.table("gabece/G2.txt", h=TRUE)
attach(Maths)
table(G2)

Sciences <- read.table("gabece/G3.txt", h=TRUE)
attach(Sciences)
```

```r
table(G3)

SES <- read.table("gabece/G4.txt", h=TRUE)
attach(SES)
table(G4)

# Cleaning ==========================================
# Variables including cases not presenting missing data -1, -2, etc..

genderA=gender[okData=="ok"]
yearA=year[okData=="ok"]
regionA=region[okData=="ok"]
G1A=G1[okData=="ok"]
G2A=G2[okData=="ok"]
G3A=G3[okData=="ok"]
G4A=G4[okData=="ok"]

# =======================================
# yearly Analysis
#In the example below, you may change 2012
#by any year in (2012, ...,2017) and
#proceed to the analysis

yearOfStudy=2012
regionOfStudy="6"

genderAY=genderA[yearA==yearOfStudy]
regionAY=regionA[yearA==yearOfStudy]
G1AY=G1A[yearA==yearOfStudy]
G2AY=G2A[yearA==yearOfStudy]
G3AY=G3A[yearA==yearOfStudy]
G4AY=G4A[yearA==yearOfStudy]

# study of region
genderAR=genderA[regionA==regionOfStudy]
G1AR=G1A[regionA==regionOfStudy]
G2AR=G2A[regionA==regionOfStudy]
G3AR=G3A[regionA==regionOfStudy]
G4AR=G4A[regionA==regionOfStudy]

# study of region and year
```

```
genderAYR=genderA[(regionA==regionOfStudy) & (yearA==yearOfStudy)]
G1AYR=G1A[(regionA==regionOfStudy) & (yearA==yearOfStudy)]
G2AYR=G2A[(regionA==regionOfStudy) & (yearA==yearOfStudy)]
G3AYR=G3A[(regionA==regionOfStudy) & (yearA==yearOfStudy)]
G4AYR=G4A[(regionA==regionOfStudy) & (yearA==yearOfStudy)]
length(G4AYR)

#proceed to your analysis
```